\documentclass[conference]{IEEEtran}
\IEEEoverridecommandlockouts

\usepackage{amsmath,amssymb,amsfonts}
\usepackage{textcomp}
\usepackage[OT1]{fontenc}
\usepackage[colorlinks]{hyperref}
\usepackage{tikz}

\usepackage[style=ieee]{biblatex}
\newcommand{\Enc}{\Call{Enc}{}}

\newcommand{\Sign}{\Call{Sign}{}}

\usepackage[
	noEnd,
	endLComment={},
]{algpseudocodex}

\newcommand{\function}[3]{
	\Function {#1} {#2}
	#3
	\EndFunction
	\smallskip
}

\RequirePackage{mathtools}
\newcommand{\ceq}{\coloneqq}

\newcommand{\Z}{\mathbb Z}

\title{Poster: Committee Moderation on Encrypted Messaging Platforms\thanks{This work was supported by NSF grant 1814753.}}

\author{
    \IEEEauthorblockN{Alistair Pattison}
    \IEEEauthorblockA{
        \textit{Carleton College, University of Minnesota} \\
        pattisona@carleton.edu
    } \and
    \IEEEauthorblockN{Nicholas Hopper}
    \IEEEauthorblockA{
        \textit{University of Minnesota} \\
        hoppernj@umn.edu
    }
}

\begin{document}

\maketitle

\section{Introduction}

Over the past decade, the increased prevalence of mobile computing and a growing desire for privacy have lead to a surge in the use of encrypted messaging services like WhatsApp, Facebook Messenger, and Signal.
The deniability, anonymity, and security provided by these services are crucial to their widespread adoption, but by construction, these properties make it impossible to hold users accountable for the messages they send.
With no accountability, these platforms are ripe for abuse, and WhatsApp group chats have been used to spread misinformation that has influenced elections \cite{pbs-brazil-whatsapp,oxford-computational-propoganda-report} and even incited the murder of a woman in India \cite{nytimes-whatsapp-india-murder}.
With no way of identifying or verifying the senders of these messages, little can be done.

Previous works \cite{hecate, tglmr} have attempted to find a middle ground between accountability and privacy by allowing a moderator to verify the original sender of a message if the message is reported; if not reported, messages maintain all security guarantees.
However, these works concentrate all responsibility  for determining if a message requires moderation to a single party.
This is undesirable.
Using primitives from threshold cryptography, this work extends the message-reporting protocol Hecate \cite{hecate} to a setting in which consensus among a group of moderators is required to reveal and verify the identity of a message's sender.

\section{Previous Works}

An obvious road to accountability is to require users to sign their messages, but this completely destroys deniability, which is an important feature in many use cases.
\textcite{tglmr} preserve deniability by using a cryptographic primitive that allows signatures to be verified only by one person who is chosen at the time of signing.
By making the designated verifier a trusted third party (for example, law enforcement or a school principal) and attaching to each message a zero-knowledge proof that the signature is valid, users can be confident that abusive messengers can be held accountable for the messages they send while still preserving deniability against everyone else.
However, this protocol uses heavy crypto machinery and is quite expensive.

More recent work by \textcite{hecate} introduced a protocol called Hecate that provides the same security guarantees as \textcite{tglmr} but is significantly cheaper in terms of the number of invocations of cryptographic primitives.
It works as follows:
In advance of sending any messages, users request ``tokens'' from a moderator containing (among other things) an encryption of the message-sender's identity,
\begin{equation}
	x_1 \ceq \Enc_{mod}(id_{src}),
\end{equation}
and a single-use ephemeral key pair $(pk_{eph}, sk_{eph})$.
The token also includes a signature $\sigma_{mod}$ made with the moderators private key that binds the key pair to $x_1$.

When a user sends a message, he consumes a token and attaches the signature
\begin{equation}
	\sigma_{src} = \Sign_{sk_{eph}}(x_2)
	\quad \text{where} \quad
	x_2 \ceq x_1 \oplus H(m)
\end{equation}
to the message along with the original moderator token.
The signature binds $x_1$ to the sent message, and this metadata is carried with the message throughout its entire forwarding chain.
If the message is ever reported, the moderator decrypts $x_1$ with her private key to obtain the original sender's identity.
To everyone else, the token provides no information.

We direct readers to \textcite{hecate} for a richer description of the protocol and its properties.

\section{Our Protocol}

Our modified protocol retains the same general flow of token-issuing and message reporting from Hecate \cite{hecate}, but modifies the process by which $x_1$ is created and decrypted.
We call our new protocol Cerberus (the name of the multi-headed dog that guards the river Styx) as a nod to the multiple moderators in the protocol and the greek name of the original Hecate protocol.

In Cerberus, there are $n$ moderators, and $k$ of them must cooperate to recover $id_{src}$ from $x_1$.
(These $k$ and $n$ values are tunable protocol parameters.)
The token-creation process is described below using Elgamal threshold encryption and the FROST \cite{frost} signature algorithm, although different threshold schemes could be substituted. In \Call{CreateToken}{}, $G$ is a (secure) group of order $q$ with generator $g$, the moderators' public encryption key is $pk_{mod}$, and the corresponding private encryption key is divided into shares $s_1, \ldots, s_n$ using a Shamir secret-sharing scheme \cite{shamir-secret-sharing}.
The token generated by \Call{CreateToken}{} is identical to a token in the Hecate protocol \cite{hecate}, and the message is processed as is described in that paper.

\smallskip

\begin{algorithmic}[1]
	\function{CreateToken}{$id_{src}$}{
		\State Generate $r \gets_\$ \Z_q$ and an an ephemeral keypair $(pk_{eph}, sk_{eph})$.
		\State Compute $x_1 \ceq \left(g^r, \ id_{src} \oplus H(pk_{mod}^r)\right)$.
		\State Package $x_1$ into a token and send a signature request to each moderator along with the randomness $r$.
		\State Each moderator verifies that $x_1$ is a valid encryption of $id_{src}$ with the provided randomness and returns their signature share on the token.
		\State Once sufficient responses are received, combine the signature shares into a valid Schnorr signature $\sigma_{mod}$.
	}
\end{algorithmic}

\smallskip

To report a message, a user sends out requests to every moderator, each of whom decides individually whether or not to respond with a decryption share $d_i$ serving as a vote that the message should be acted upon.
If more than $k$ decryption shares are received, i.e., if more than $k$ moderators believe that the message requires moderation, then one can recover the identity of the sender, $id_{src}$. If there are insufficient responses, $id_{src}$ remains hidden.
This process of ``voting'' adresses the question posed in \cite{hecate} of how to handle reported messages that are not necessarily abusive or misinfirmative. A formal description is as follows:

\smallskip

\begin{algorithmic}[1]
	\function{HandleReport}{$m$, $T = (x_1, \ldots )$}{
	\State A user sends a request to all $n$ moderators containing the reported message $m$ and its attatched token $T$ containing the encrypted id, $x_1 = (c_1, c_2)$.
	\State Each moderator verifies the token and---if she believes that the message requires intervention---responds with the decryption share $d_i \ceq c_1^{s_i}$.
	\State If $k$ decryption shares are received, they can be combined with appropriate Lagrange coefficients $\lambda_i \ceq \prod_{j \neq i} \frac{j}{j - i}$ to obtain
	\begin{equation}
		id_{src}
		= H \left(\prod_i d_i^{\lambda_i}\right) \oplus c_2.
	\end{equation}
	}
\end{algorithmic}

\section{Benchmarks}

\textcite{hecate} includes an implementation and benchmark of the whole message cycle, so we focus on the modified portions of the protocol: token creation and message reporting.
We implement these steps in Rust and run each party in a separate Linux container communicating over HTTP. Source code and more details are available on github \cite{cerberus-github}.
Cerberus is much slower to create tokens than Hecate (on the order of 50x), but this isn't a huge surprise: the benchmarks were run on slower hardware, and the distributed nature of this protocol involves a multitude of verification, serialization, and communication steps that are not required in Hecate.
With that said, the operational costs of implementing a protocol such as this are still \textit{well} within the budgetary constraints of a large company like Facebook. Extending the analysis in \cite{hecate}, we estimate the total cost of running all necessary servers to be under \$100 a day for the entirety of WhatsApp.

\begin{table}[!t]
	\renewcommand{\arraystretch}{1.3}
	\caption{Mean end-to-end runtimes (ms).}
	\label{tab:mean-times}
	\centering\begin{tabular}{c | c c c}
		$n$ & 3     & 5     & 7    \\
		\hline \hline
		Token creation (per token)
		    & 1.86  & 3.06  & 4.61 \\
		Report handling (per report)
		    & 0.626 & 0.904 & 1.18
	\end{tabular}
\end{table}

\begin{figure}[!t]
	\centering
	\footnotesize
	\input{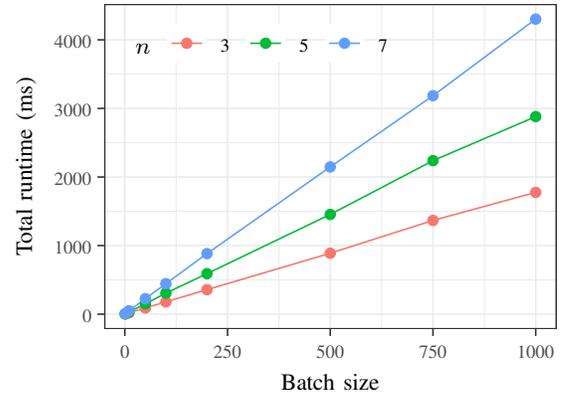}
	\caption{End-to-end token creation times as a function of batch size and number of moderators. Times include all rounds of communication.}
	\label{fig:token-creation}
\end{figure}


\printbibliography

\end{document}